\documentclass[conference,9pt]{IEEEtran}
\IEEEoverridecommandlockouts
 
\usepackage{cite}
\usepackage{amsmath,amssymb,amsfonts}
\usepackage{algorithmic}
\usepackage{graphicx}
\usepackage{textcomp}
\usepackage{xcolor}
\def\BibTeX{{\rm B\kern-.05em{\sc i\kern-.025em b}\kern-.08em
    T\kern-.1667em\lower.7ex\hbox{E}\kern-.125emX}}
   
\begin{document}

\title{Memcomputing for Accelerated Optimization}

\author{\IEEEauthorblockN{John Aiken}
\IEEEauthorblockA{
\textit{MemComputing, Inc.}
San Diego, California, 92121 \\
jaiken@memcpu.com}
\and
\IEEEauthorblockN{ Fabio L. Traversa}
\IEEEauthorblockA{
\textit{MemComputing, Inc.}
San Diego, California, 92121 \\
ftraversa@memcpu.com}
\and

}

\maketitle

\begin{abstract}
In this work, we introduce the concept of an entirely new circuit architecture based on the novel, physics-inspired computing paradigm: Memcomputing. In particular, we focus on digital memcomputing machines (DMMs) that can be designed leveraging properties of non-linear dynamical systems; ultimate descriptors of electronic circuits. The working principle of these systems relies on the ability of currents and voltages of the circuit to self-organize in order to satisfy mathematical relations. In particular for this work, we discuss self-organizing gates, namely Self-Organizing Algebraic Gates (SOAGs), aimed to solve linear inequalities and therefore used to solve optimization problems in Integer Linear Programming (ILP) format. Unlike conventional I\textbackslash O gates, SOAGs are “terminal-agnostic”, meaning each terminal handles a superposition of input and output signals. When appropriately assembled to represent a given ILP problem, the corresponding self-organizing circuit converges to the equilibria that express the solutions to the problem at hand. Because DMM’s components are non-quantum, the ordinary differential equations describing it can be efficiently simulated on our modern computers in software, as well as be built in hardware with off-of-the-shelf technology. As an example, we show the performance of this novel approach implemented as Software as a Service (MemCPU XPC) to address an ILP problem. Compared to today’s best solution found using a world renowned commercial solver, MemCPU XPC brings the time to solution down from 23 hours to less than 2 minutes. 

\end{abstract}


\subsubsection*{\bf{INTRODUCTION}}
The never-ending search for greater compute power and efficiency is propelling both the industrial and scientific community to explore new and unique computational paradigms to overcome the current bottlenecks associated with modern computing techniques. Memcomputing (which stands for computing with and in memory) \cite{UMM,DMM2,Di_Ventra2018} is one such technology, and has shown to deliver significant improvements to the solutions of hard combinatorial/optimization problems compared to traditional algorithmic approaches \cite{Sheldon2019,Traversa2018,Traversa2019,Sheldon2019a,Traversa2018a,AcceleratingDL}. 
In this work, we firstly describe the circuit architecture of memcomputing machines specifically designed to solve the linear inequalities that represent the building blocks of a general Integer Linear Programming (ILP) problem~\cite{Schrijver1998}. This architecture, previously introduced by one of the authors (FT) in \cite{Traversa2018a},  dynamically self-organizes to reach an equilibrium that is a representation of a solution for the linear inequality. We then show how the emulation of these circuits through our MemCPU™ ILP solver finds the solution to a hard benchmark problem from the Mixed Integer Programming Library 2017 \cite{MIPLIB2010}, and compare it to state of the art commercial solver Gurobi.

	\begin{figure}
\centerline{\includegraphics[width=1\columnwidth]{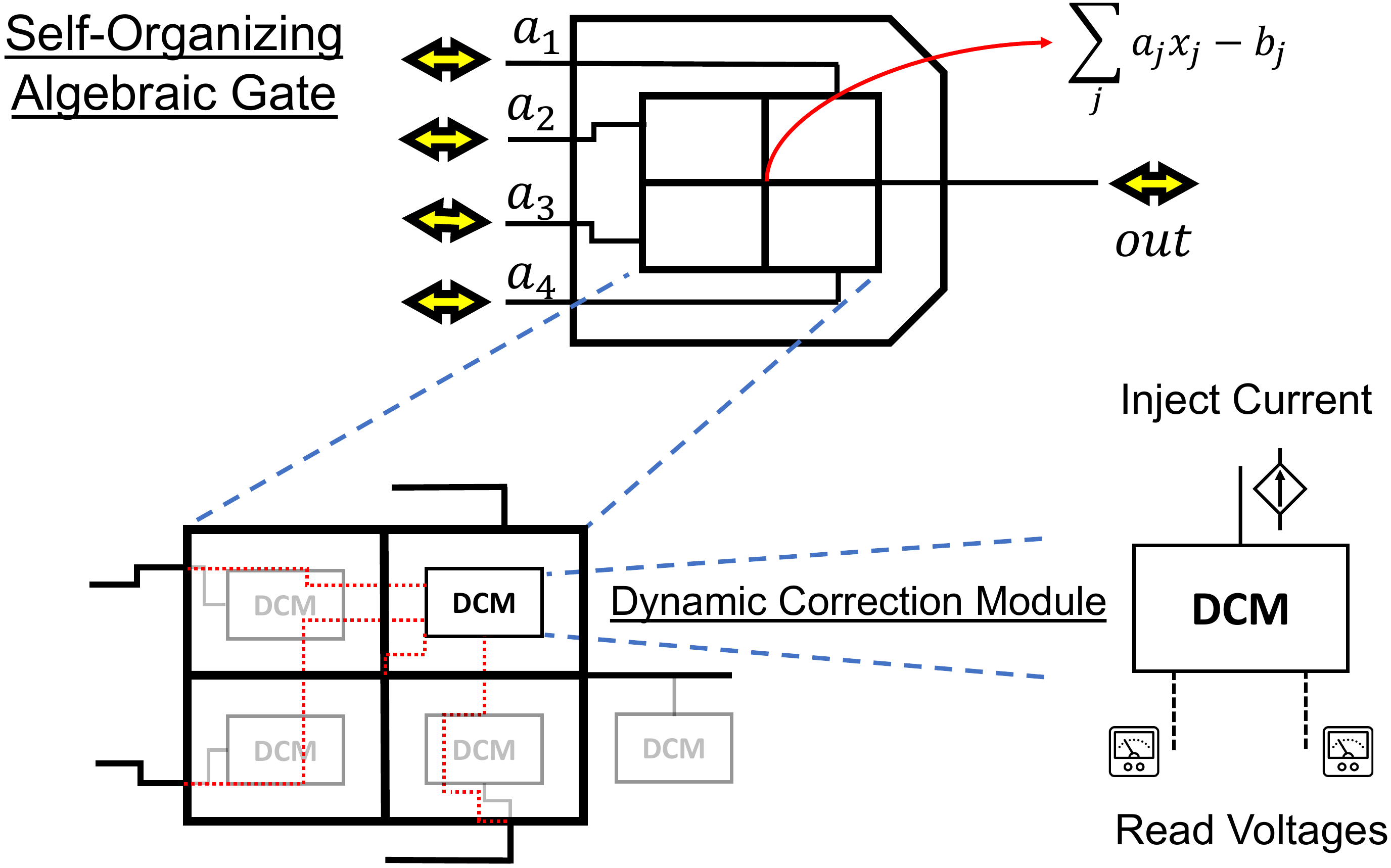}}
    \caption{Sketch of a Self-Organizing Algebraic Gate (SOAG). All terminals allow a superposition of incoming and outgoing signals from the surrounding circuit. The central unit processes the signals in order to satisfy a linear algebraic relation consistent with the requirement of the “out” terminal. The self-organization is enforced by the Dynamic Correction Modules that read voltages from all terminals and inject an appropriate current as long as the algebraic relation is not satisfied.}\label{fig_SOAG}
\end{figure}

\subsubsection*{{\bf SELF-ORGANIZING ALGEBRAIC GATES}}

Memcomputing’s approach to addressing ILP problems is based on the novel concept of Self-Organizing Algebraic Gates (SOAGs) \cite{Traversa2018a} (see Fig.~\ref{fig_SOAG}). SOAGs represent an entirely new circuit design developed at MemComputing, Inc. and was inspired by the previous work on Self-Organizing Logic Gates (SOLGs) \cite{DMM2,Di_Ventra2018}. Both SOAGs and SOLGs represent the foundation for practical realizations of Universal MemComputing Machines (UMMs) \cite{UMM,DMM2,Di_Ventra2018}, specifically their digital, thus scalable sub-class: Digital Memcomputing Machines (DMMs) \cite{DMM2,Di_Ventra2018}. 

SOAGs share the same principles and scalability advantages as SOLGs, such as long-range order and topological robustness \cite{topo,no-chaos}, while also avoiding persistent chaotic and oscillatory behavior \cite{no-chaos,noperiod} which proves to be very effective when addressing a variety of combinatorial optimization problems such as maximum satisfiability (MAXSAT) \cite{Sheldon2019,Traversa2018}, quadratic unconstrained binary optimization (QUBO) and spinglass problems \cite{Sheldon2019a}, and training of deep networks \cite{AcceleratingDL}.  

However, whereas SOLGs self-organize toward a boolean relation, SOAGs are designed to self-organize to satisfy linear inequalities between boolean variables (see Fig.~\ref{fig_SOAG}). Since an ILP problem is the issue of finding an assignment to integer variables subject to a collection of constraints in the form of linear inequalities \cite{Schrijver1998}, SOAGs can be used to represent the constraints of the ILP problem. Therefore, when connecting together the terminals of SOAGs sharing the same variables, we assemble a Self-Organizing Algebraic Circuit (SOAC) that embeds the ILP problem. When properly designed, the dynamics of the SOAC induces the collective self-organization of all the SOAGs in the circuit towards an equilibrium of currents and voltages in which all SOAGs satisfy their algebraic relations at once; otherwise no equilibrium is reached~\cite{Traversa2018a}. Therefore, each equilibrium represents a feasible solution for the ILP problem.


\subsubsection*{{\bf DEMONSTRATED PERFORMANCE }}

\begin{figure}
\centerline{\includegraphics[width=.98\columnwidth]{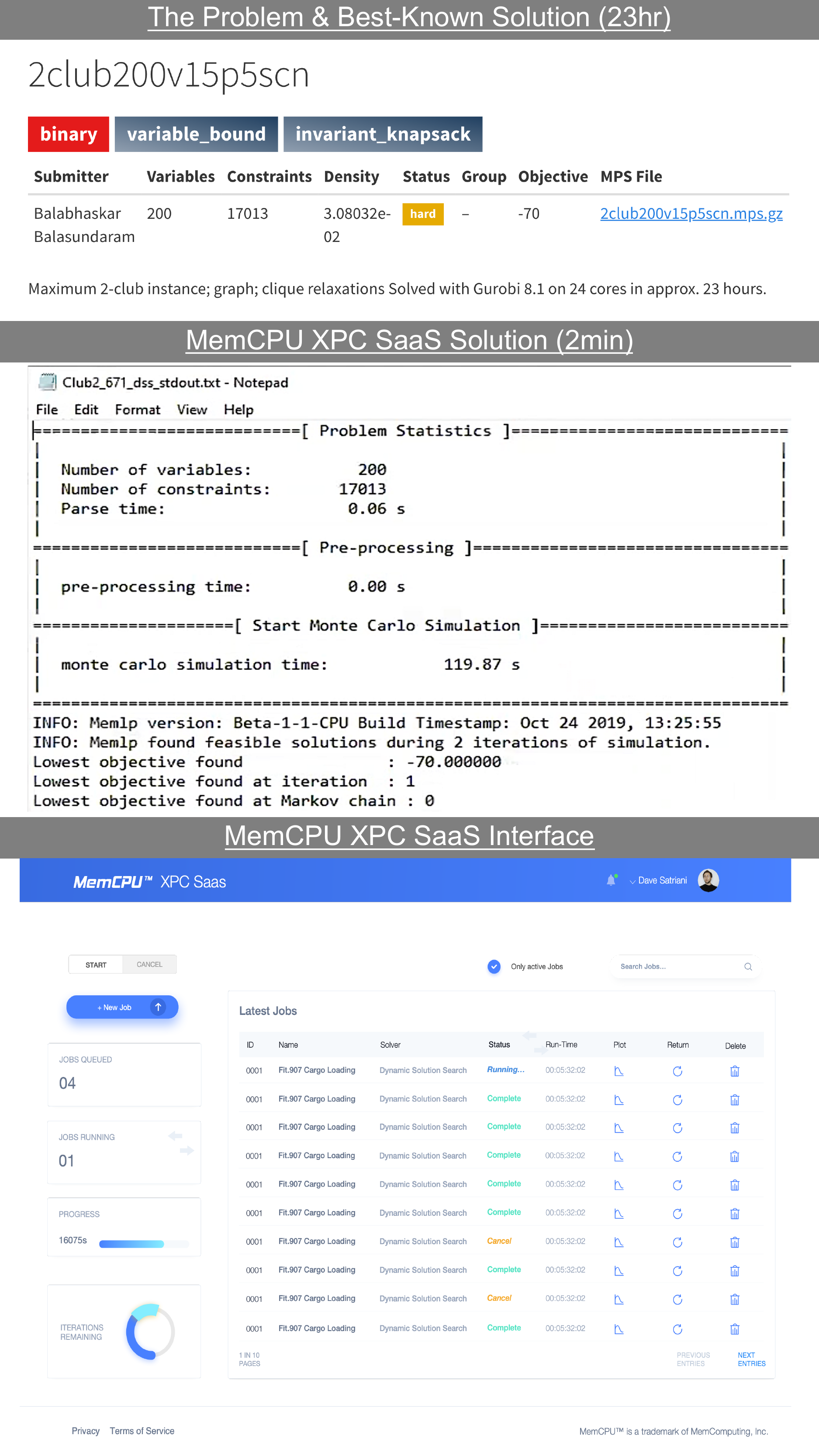}}
    \caption{Top section shows the details of  the Club2 problem from the MIPLIB website and the results of the best known solution. Middle section represents the output found using the MemCPU ILP solver when solving the Club2 problem. Last section displays the interface of the MemCPU XPC SaaS.}\label{fig_HCF2}
\end{figure}

As a benchmark, we consider a problem taken from the Mixed Integer Programming Library (MIPLIB) 2017 ~\cite{MIPLIB2010}. MIPLIB is currently managed by the Zuse Institute of Berlin, and has been since the 1990's, and collects problems from industry and academia and classifies them based on the performance of commercial solvers. For example, the Club2 problem belongs to a class of problems classified as “hard”, meaning that it takes best in class commercial solvers no less than 1 hour to find the global optimum~\cite{MIPLIB2010}. In Fig.~\ref{fig_HCF2} top section, there is an extract from the MIPLIB website reporting that the global optimum was found in 2018 using Gurobi 8.1 after running for 23 hours on a 24 core processor. Gurobi is a prominent commercial solver for mixed-integer programming (MIP), and employs sophisticated algorithms and heuristics in order to further accelerate their branch-and-bound procedure. The result is a collection of state-of-the-art algorithms, solution strategies and optimization toward the solution of MIP problems.

Here we stress the major difference between our MemCPU ILP solver and Gurobi 8.1. The MemCPU ILP solver solves differential equations of a physical system (the SOAC) that represents the original ILP problem. Gurobi, instead, represents a sophisticated but still traditional (combinatorial) algorithmic approach. In other words, the memcomputing approach first transforms the original optimization problem into a physics problem, and then simulates the dynamics of such a physical system, while maintaining the digital structure of inputs and outputs \cite{Di_Ventra2018}.

In order to compare the performance of the MemCPU ILP solver to Gurobi 8.1, we ran the Club2 problem using the MemCPU™ Extreme Performance Computing (XPC) Software-as-a-Service (SaaS) platform; our online platform that leverages virtual machines through the cloud to run the MemCPU ILP solver. Our solver then processed the problem, generating an SOAC that exactly represented the ILP problem while running on a CPU virtual machine with 36 cores (see Fig. 2 middle section). 

From the overall comparisons, we can see that the approach taken by the MemCPU ILP solver found the global optimum (-70) in just 120 seconds while Gurobi took 23 hours.

\subsubsection*{{\bf CONCLUSIONS}}

In this work, we have discussed an entirely new circuit design composed of self-organizing gates. When assembled together, they create a circuit, which collectively self organizes to satisfy algebraic relations; thus showing how to employ scalable Digital Memcomputing Machines to address and solve the complex, yet very important class of integer linear programming problems. 

We have simulated the corresponding equations of motion of these circuits in software accessed through the MemCPU XPC SaaS,  which employed the MemCPU ILP solver to find the global optimum to the Club2 problem, a hard benchmark problem from the MIPLIB 2017 library. We compared the results of our solver to the best known solution, found by Gurobi 8.1. The MemCPU ILP solver proved to be much more efficient in converging to the global optimum, finding it in just 120 seconds versus Gurobi’s 23 hour compute time. 

Because memcomputing machines leverage non-quantum systems, they can easily be implemented in hardware using standard electronic components for these, and other challenging combinatorial optimization problems.

\bibliographystyle{naturemag}
\bibliography{SUSYref}

\end{document}